# Measurement of the energy distribution of electrons escaping minimum-B ECR plasmas

I.Izotov[1], O.Tarvainen[2], V. Skalyga[1], D. Mansfeld[1], T. Kalvas[2], H. Koivisto[2], R. Kronholm[2]
1 Institute of applied physics of Russian academy of sciences, Nizhny Novgorod, Russia
2 University of Jyvaskyla, Finland

## Abstract

The measurement of the electron energy distribution (EED) of electrons escaping axially from a minimum-B electron cyclotron resonance ion source (ECRIS) is reported. The experimental data were recorded with a room-temperature 14 GHz ECRIS at the JYFL accelerator laboratory. The electrons escaping through the extraction mirror of the ion source were detected with a secondary electron amplifier placed downstream from a dipole magnet serving as an electron spectrometer with 500 eV resolution. It was discovered that the EED in the range of 5 - 250 keV is strongly non-Maxwellian and exhibits several local maxima below 20 keV energy. It was observed that the most influential ion source operating parameter on the EED is the magnetic field strength, which affected the EED predominantly at energies less than 100 keV. The effects of the microwave power and frequency, ranging from 100 to 600 W and 11 to 14 GHz respectively, on the EED were found to be less significant. The presented technique and experiments enable the comparison between direct measurement of the EED and results derived from bremsstrahlung diagnostics, the latter being severely complicated by the non-Maxwellian nature of the EED reported here. The role of RF pitch angle scattering on electron losses and the relation between the EED of the axially escaping electrons and the EED of the confined electrons are discussed.

## Introduction

Microwave discharges are widely used as sources of positive and negative ions and ion beams as well as in plasma technology, including thin-film deposition, plasma etching, surface ion treatment and sputtering . A significant fraction of these discharges operate at low gas pressure with the electrons confined magnetically and being heated by microwaves under the electron cyclotron resonance (ECR) condition where the electron gyrofrequency equals the microwave frequency, therefore enabling an efficient electron heating and subsequent production of high charge state ions in ionizing collisions between the electrons and neutrals / ions. ECR ion sources (ECRIS) have been essential in accelerator based nuclear physics research and applications over the past 40 years. They are extensively used in a wide range of large-scale accelerator facilities for the production of highly charged ion beams of stable and radioactive elements, from Hydrogen up to Uranium.

One of the main peculiarities of ECR heating is that electrons gain mainly (depending on their energy [1-3]) transverse energy when interacting with the microwave electric field, thus making the electron velocity distribution function (EVDF) of ECRIS plasmas strongly anisotropic. The resulting electron energy distribution function (EEDF) is believed to consist of (at least) three main components: cold electrons with an average energy $<E_{e,cold}>$ of 10–100 eV, warm electrons with $<E_{e,warm}>$ of 1–10 keV and hot electrons with $<E_{e,hot}>$ from 10 keV up to 1 MeV [4-6]. Together with particle densities the EEDF defines the volumetric reaction rates in ECRIS plasmas and affects the growth and damping rates of non-linear processes. In particular, it has been shown that kinetic instabilities originating from the anisotropy of the EVDF and non-equilibrium EEDF cause unacceptable periodic oscillations of the extracted ion beam due to a loss of the ion confinement at periodic intervals [7]. Thus, knowledge on the EEDF and its control is crucial for optimization of ECRIS performances and benchmarking PIC simulation codes.

In the case of rarefied plasma electrons are gaining mostly transverse energy, which then becomes higher than the ionization potential and longitudinal kinetic energy [8]. Under such condition, the interaction with the microwaves determines both the acceleration of the electrons produced as a result of the stepwise ionization and their scattering to the loss cone (i.e. RF-scattering [1,2]). When electrons interact with a monochromatic electromagnetic wave in the ECR region, they are heated stochastically (the Fermi acceleration regime). The ensemble of electrons diffuses in energy determined by the quasilinear diffusion coefficient [1]. When the transverse electron energy reaches a certain value, the perturbation of the phase shift of the cyclotron interaction, which arises as a result of the nonadiabatic nature of the motion in the ECR zone, becomes negligible. This leads to the destruction of the Fermi acceleration regime, and the particle energy begins to oscillate around a certain cut-off value and the electron motion becomes superadiabatic. Under such condition, the EEDF forms a "quasilinear plateau" in the resonance region of the momentum space. In this case, the electrons can not gain energies exceeding a certain value and, thus, at this stage of the discharge the ability of the plasma to absorb the microwave energy is limited as the energy is being absorbed only by the growth of the plasma density, not by changing the shape of the EEDF. Such conditions are usually realized at the initial stage of the discharge, which explains the high reflection coefficient associated with the plasma breakdown as observed e.g. in [9]. The EEDF with the quasilinear plateau formed at the beginning of the ECR discharge is also responsible for the Preglow effect [10-12]. When collisions are introduced at higher plasma density and/or the wave is not strictly monochromatic, the superadiabatic EEDF partly collapses since the phase of the electron-wave interaction remains random for a part of electrons, which enables "overheating" of the electrons up to 1 MeV energies as detected through bremsstrahlung diagnostic [13]. Furthermore, collisions and collision-like processes alter the electron diffusion lines in momentum space and, thus, affect the EEDF through complex interactions. The resulting energy distribution is considered to have a sophisticated shape, being strongly non-Maxwellian.

Indirect characterization of the EEDF based on plasma bremsstrahlung spectroscopy is a simple, well-known and widely used method of probing the effect of different ion source operational parameters on the high-energy photon emission spectrum at energies above 1 keV (for a recent study see e.g. [14] and references therein). However, the technique does not give neither qualitative nor quantitative information on EEDF without complicated deconvolution of the spectrum, often compromised by the experimental geometry (plasma vs. wall bremsstrahlung) and the inherent sensitivity to assumptions [15]. Measurement of the plasma bremsstrahlung is therefore mostly used for determining a "spectral temperature", which has little value for precise analysis of the EEDF and nonlinear plasma-wave interactions, but can be considered as a qualitative indicator of the plasma energy content as a function of operational parameters. Also, the maximum energy of the electrons can be derived from the "endpoint energy" of the bremsstrahlung spectrum. However, the EEDF can be measured directly. One of the simplest method is applying Langmuir probe diagnostics and subsequent analysis of the I-V characteristics with Druyvesteyn theory [16]. However, even the modified theory, being successful for measuring EEDF in a low-temperature microwave-heated plasmas [17] of singly charged ions, is inapplicable for ECRIS of multicharged ions. This is due to the invasive nature of the probe, which perturbs the plasma equilibrium and distorts the EEDF. Furthermore, the Langmuir probe techniques is inapplicable for measuring electron energies in keV - MeV range especially in strong, spatially varying magnetic fields.

The present work reports the first (to our knowledge) direct measurements of the energy distribution function of electrons escaping the magnetic confinement of conventional minimum-B ECRIS in stable CW operation i.e. in the absence of kinetic instabilities. It is emphasized that the EEDF of the confined electrons in the magnetic trap and the EEDF of the escaping electrons might be different. However, the following considerations allow to argue that the EEDF of the escaping electrons reflects the EEDF of the confined electrons (at least in terms of parametric dependencies).

A resonant interaction between an electron and the microwave increases both transversal and longitudinal momentum [1, 2]. The increment of the longitudinal momentum increases with the electron energy, thus energetic electrons diffuse to the loss cone in the momentum space and leave the magnetic trap in a process referred as RF-scattering. This model makes it possible to qualitatively explain how energetic electrons leave the trap. Unfortunately even the relativistic model cannot explain all peculiarities of the EEDF found so far in the experiments. Measurements reported in [1, 2, 18] and in the present paper (see section "Experimental results") have demonstrated that the energy distribution of the electrons leaving the trap spans over a wide range of energies. On the contrary the simplified relativistic model [1, 2] predicts that electrons leave the trap with a fixed energy, determined by the ratio between the microwave frequency and the electron cyclotron frequency at the mirror point. This result is related to the simplifying assumptions of the model, stating that the plane electromagnetic wave interacting with the electrons propagates along the magnetic field and has relatively weak amplitude and small vacuum wave number. In practice the observed spread of the electron energy can be caused by collisions, perturbations and damping of the incident electromagnetic wave, and by the fact that the heating wave is not monochromatic. RF-induced scattering of electrons in ECR-heated plasmas has been reported e.g. in Refs. [1, 2]. Recent observations have demonstrated that the RF-scattering contributes significantly to electron losses in the range of 20 - 570 keV [18, 19]. This enables characterizing the EEDF inside the trap by measuring the EEDF of the electrons escaping the confinement, yet the relative importance of this mechanism on the total electron losses and the energy-dependence of its efficiency remains unknown.

# Experimental setup

The experimental data were taken with the JYFL 14 GHz ECRIS. The source uses an Nd-Fe-B permanent magnet sextupole arrangement and two solenoid coils. The superposition of the solenoid and sextupole fields forms a minimum-B structure for confinement of the plasma. The strength of the permanent magnet sextupole is 1.09 T on the magnetic poles and 0.70 T between the poles, both values given at the chamber wall. The axial field strength can be varied by adjusting the solenoid currents, which affects the injection and extraction mirror ratios as well as the $B_{min}/B_{ECR}$ ratio ($B_{ECR}$=0.5 T at 14 GHz). The solenoid field configuration is best described by the values at injection ($B_{inj}$), minimum ($B_{min}$), and extraction ($B_{ext}$). For the settings corresponding to a typically used $B_{min}/B_{ECR}$ = 0.75 the values are $B_{inj}$=0.913 T, $B_{min}$=0.375 T and $B_{ext}$=1.976 T. $B_{min}/B_{ECR}$ ratio is given later for each experiment being the most convenient for describing the magnetic field strength. Plasma electrons are typically heated by 100–600 W of microwave power at 14 GHz. The source is equipped with a secondary waveguide port connected to a 10.75 - 13.75 GHz TWT amplifier with 350 - 400 W maximum power. Typical operating neutral gas pressures are in the $10^{-7}$ mbar range.

The electrons escaping the confinement were detected with a secondary electron amplifier placed in the beamline downstream from the 90 degree bending magnet used as an energy dispersive separator. The electron flux was limited by two $\phi$ = 5 mm collimators placed between the ion source and the bending magnet and yet another $\phi$ = 5 mm entrance collimator in front of the secondary electron amplifier. The experimental setup is shown schematically in Fig. 1. The polarity of the bending magnet power supply was changed from the normal operation where the magnet is used for m/q-separation of high charge state positive ions. The magnetic field deflecting the electrons was measured with a calibrated Hall-probe. The energy distribution of the electrons escaping from the confinement was then determined by ramping the field of the bending magnet and detecting the electron current from the amplifier with a picoammeter. At the given field strength of the bending magnet the apparatus detects electrons with relativistic momentum $p = \gamma m_0 V = R |e| B$, where $\gamma$ is the Lorentz factor, $m_0$ - electron rest mass, $V$ - transverse speed of electrons, $R$ - radius of curvature of particle trajectories inside the bending magnet, and $B$ - the magnetic field strength. Then, $\gamma = \sqrt{1 + \left(\frac{p}{m_0 c}\right)^2}$, and the electron energy $\varepsilon = m_o c^2 (\gamma - 1)$, where $c$ is the speed of light (all in SI). The energy resolution of the setup provided by the set of collimators is estimated to be better than 500 eV. The energy dependent transmission efficiency of the electrons leaking from the ion source through the beamline sections and the bending magnet was calculated assuming that the electron distribution at the extraction aperture is independent of energy and has a KV-distribution [20]. The first two collimators sample a fraction of the beam, which is directly proportional to the energy of the electron beam as long as the beam completely illuminates the collimators (electron energy <100 MeV). Furthermore, the energy dependent yield [21] of the secondary electrons released from the amplifier cathode was taken into account during the data analysis together with electron backscattering coefficient [22]. The power supply used for operating the bending magnet coil had a high precision and small current step (the corresponding electron energy step was <100 eV), but it was limited in maximum current, which prohibited the detection of electrons with energies of >250 keV.

The amplifier (see Fig. 1) functions by emitting secondary electrons from biased aluminum cathode and amplifying the signal by a chain of subsequent meshes before measuring the current from the grounded anode. The cathode of the secondary electron amplifier was biased negatively to -4 kV with respect to the laboratory ground and the ion source, thus prohibiting the detection of electrons with energies below 4 keV.

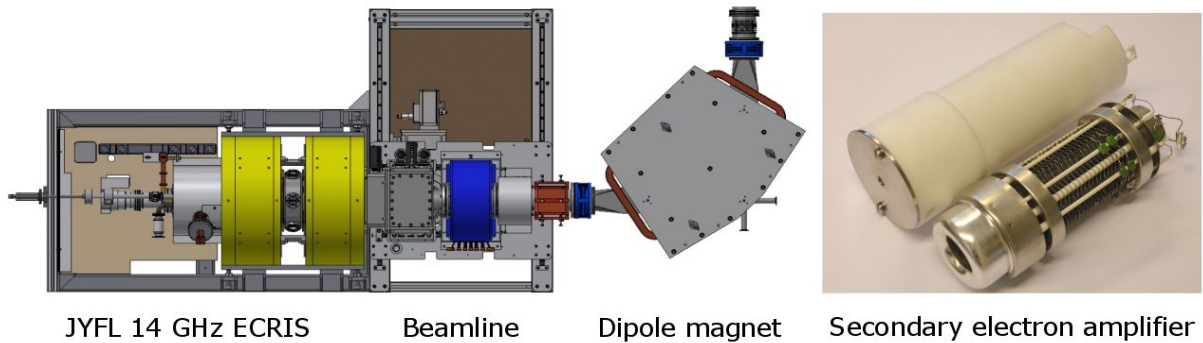

JYFL 14 GHz ECRIS    Beamline    Dipole magnet    Secondary electron amplifier

Figure 1. A schematic view of the experimental setup. From left to right: The JYFL 14 GHz ECRIS, low energy beamline with 5 mm collimators placed between the solenoid (blue) and dipole magnets, the 90 degree dipole magnet used as an electron spectrometer and the secondary electron amplifier placed at the end of the displayed beamline section following the dipole magnet. The insulating cover (white) and ϕ = 5 mm entrance collimator of the amplifier have been removed for illustration purposes to expose the amplifier chain.

The energy distribution of the escaping electrons (EED) was measured as a function of the ion source parameters e.g. microwave power, microwave frequency and (axial) magnetic field strength. The plasma chamber of the ion source and all focusing electrodes were connected to the laboratory ground throughout the experiment. This means that the detected electron flux consists of the electrons leaking from the plasma through the extraction aperture retarded only by the plasma potential of approximately 20 V [23].

## Experimental results

An example of EED obtained with the procedure described above is shown in Fig. 2. The ion source parameters were the following: 600 W of 14 GHz microwave power, $B_{min}/B_{ECR}$=0.79 and 3.5E-7 oxygen pressure (hereinafter the given value was measured without plasma). The plot is normalized to the total number of electrons i.e. the integrated signal is equal to unity. The EED has a distinct maximum at 7 keV surrounded by several subpeaks; at 15 keV the EED exhibits a noticeable drop with a transition to a Maxwellian-like[1] tail. Yet another maximum starts to appear at ~100 keV peaking above 250 keV, which was the maximum energy in our experiment, limited by the bending magnet power supply. It is emphasized that the EED shown in Fig. 2 differs from a Maxwellian one, which is often used as an assumption for the warm electron population in 1 - 100 keV range [4-6]. Furthermore, the deviation from a Maxwellian distribution questions the use of the concept of electron temperature to characterize the whole warm electron population at least when the escaping electrons are concerned.

---
[1] See the discussion section

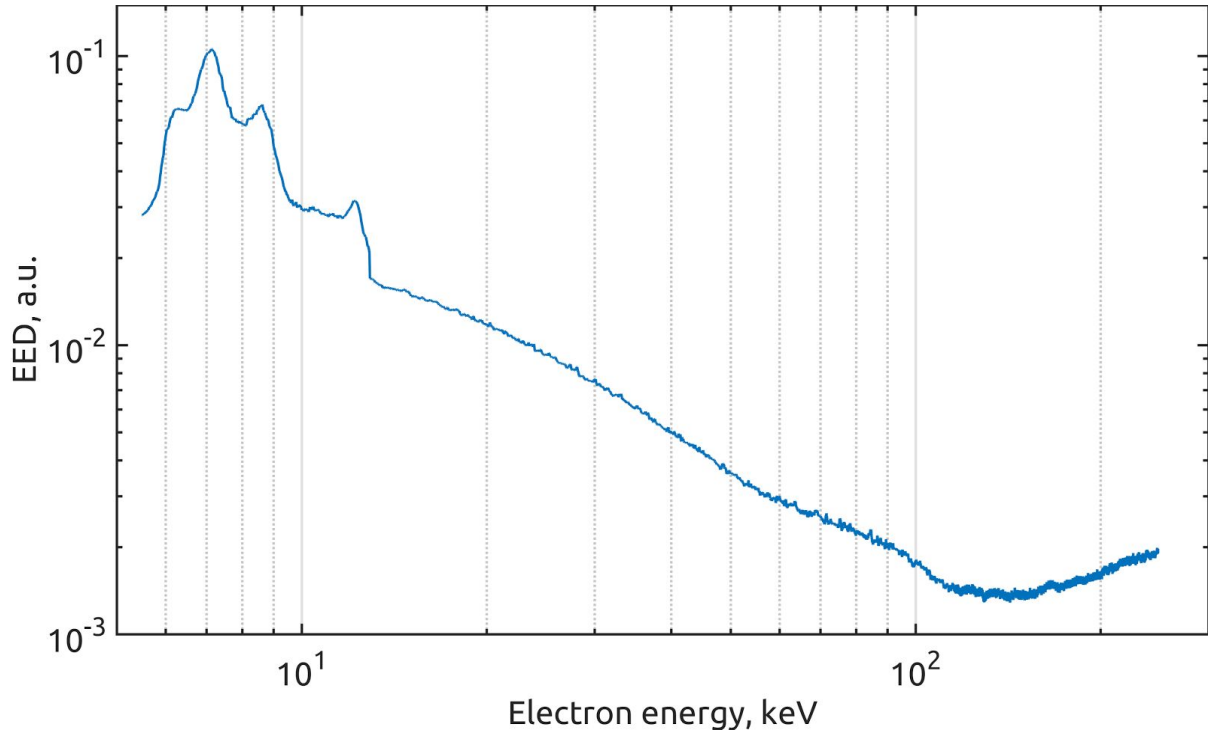

Figure 2. An example of the EED obtained with 600 W microwave power at 14 GHz, 3.5E-7 mbar oxygen pressure and $B_{min}/B_{ECR}$=0.79.

**Dependence of the EED on microwave power**

The evolution of the EED at fixed $B_{min}/B_{ECR}$=0.79 as a function of the microwave power at 14 GHz is shown in Fig. 3a. The distributions are plotted in "absolute" units, i.e. the integrated signal is proportional to the total number of detected electrons, to emphasize the dependence of the total electron flux on the injected power. Increasing the power hardly affects the position of the maxima, but rather influences only the total number of lost electrons. The distribution peaks at 7 keV despite of the heating power. Fig. 3b shows the (normalized) total number of registered electrons as a function of the microwave power.

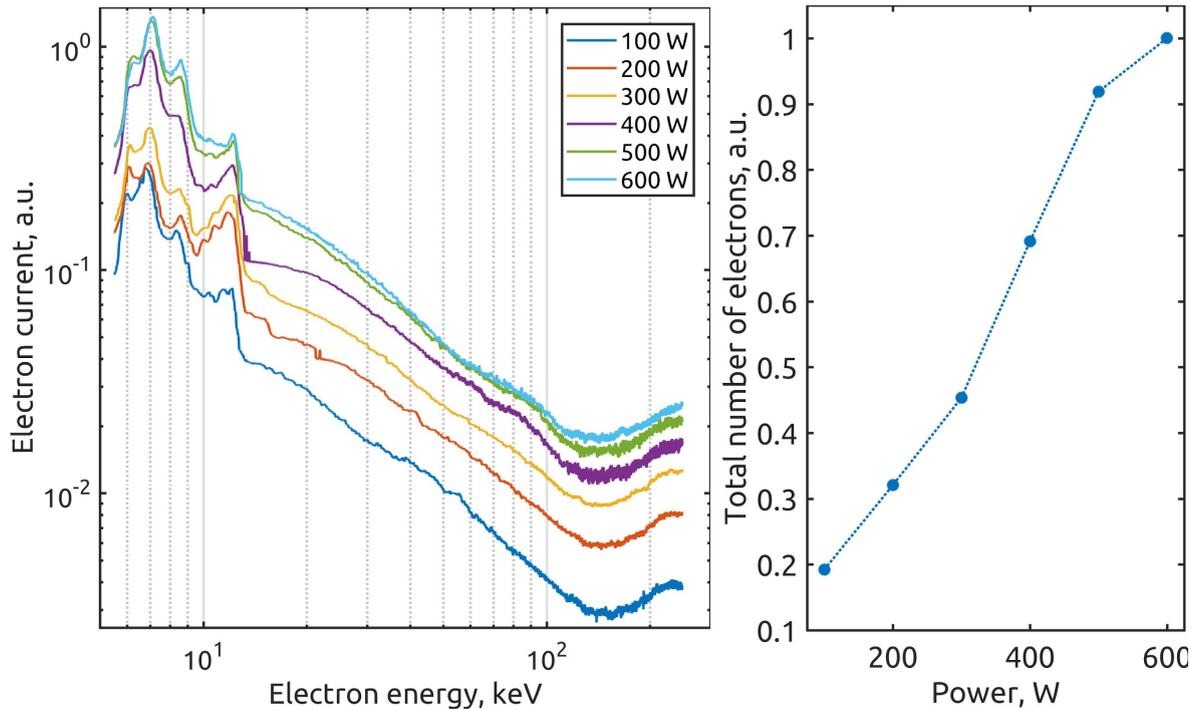

Figure 3. The EED (a) and total number of electrons (b) as a function of microwave power at 14 GHz, Bmin/BECR=0.79 and oxygen pressure of 3.5E-7 mbar.

The increase in longitudinal electron flux escaping the magnetic confinement with increasing microwave power shown in Figures 3a and 3b indicates that the plasma density and/or RF scattering rate increase with the injected power.

### Dependence of the EED on microwave frequency

It was observed that the microwave frequency did not have a pronounced influence on the shape of the EED either. The effect of the microwave frequency was probed with the TWT amplifier operating in the range of 11.0-12.4 GHz with 50 MHz step with constant power of 50 W. The operating gas was oxygen at 3.5E-7 mbar pressure and the magnetic field was kept constant at $B_{min}/B_{ECR}$=0.77. Figure 4 shows the EEDs acquired at different frequencies - they all have a similar shape and differ only in total number of electrons. The dependence of the total number of detected electrons on the microwave frequency is plotted in Fig. 5. The obvious irregularity is consistent with the frequency dependence of the ion source performance [24] and is presumably explained by the efficiency of microwave coupling including losses in the waveguide components (vacuum window, high-voltage break etc.).

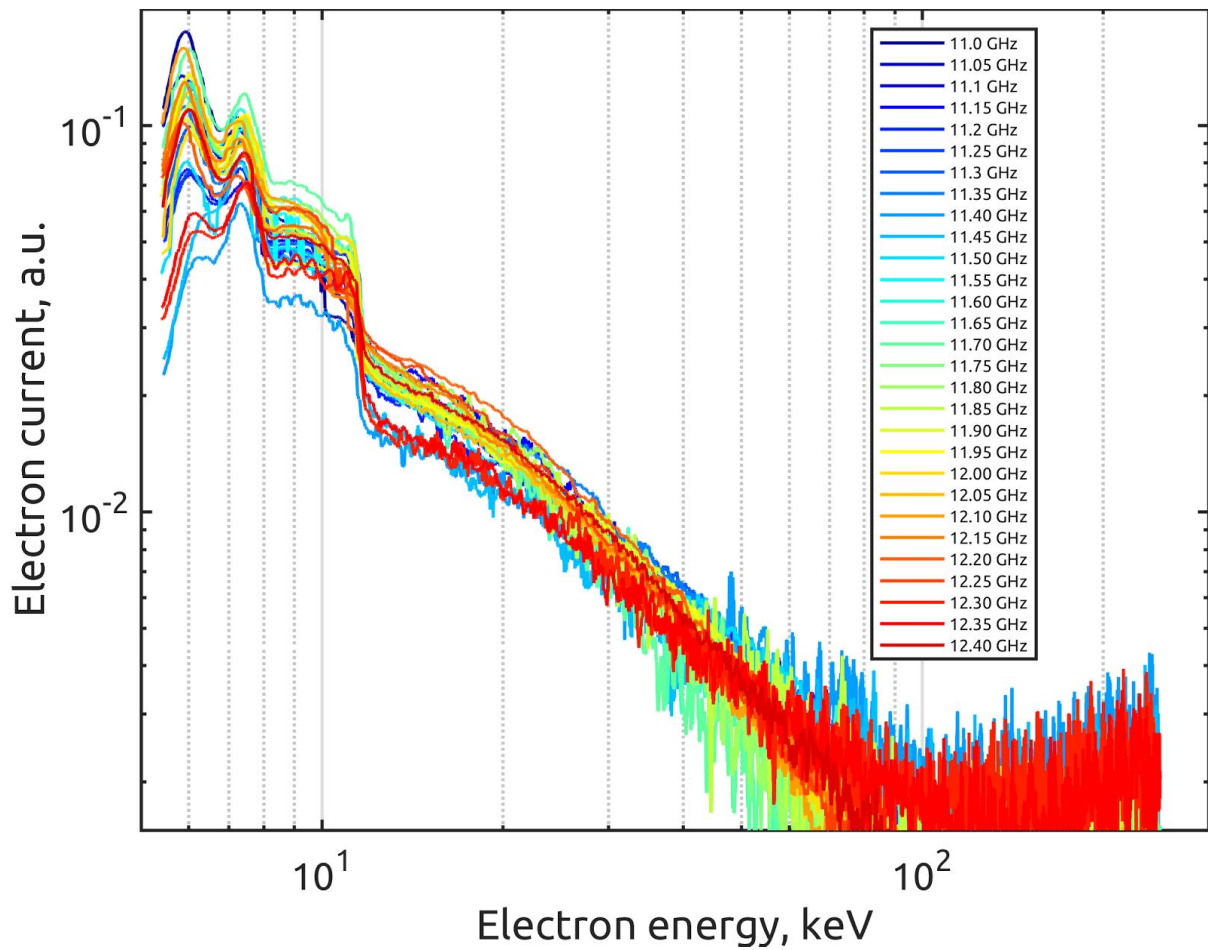

Figure 4. The effect of the microwave frequency on the EED. TWTA frequency: 11.0 - 12.4 GHz with 50 MHz, constant power of 50 W. Oxygen, 3.5E-7 mbar, $B_{min}/B_{ECR}$=0.77.

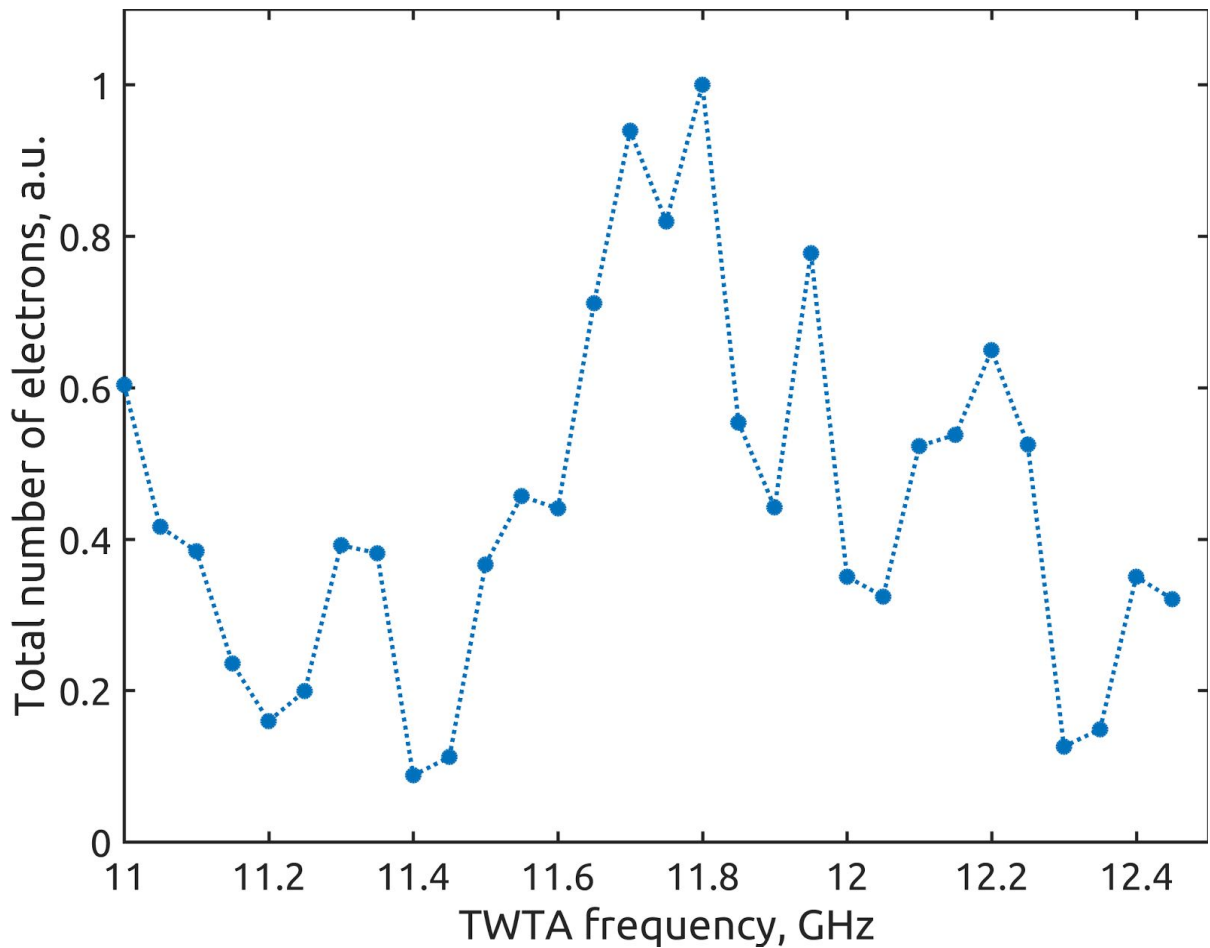

Figure 5. The total number of electrons (normalized) as a function of the microwave frequency. TWTA frequency: 11.0 - 12.4 GHz with 50 MHz, constant power of 50 W. Oxygen, 3.5E-7 mbar, $B_{min}/B_{ECR}$=0.77.

### The effect of two-frequency heating on the EED

Studies performed in the case of two-frequency heating, which is widely used method of improving ECRIS performance [25] (yet the exact mechanism remains unclear), did not reveal a significant change of the EED in comparison to single frequency heating discussed above. Figure 6 illustrates this by showing EEDs observed in the following combinations of microwave power / frequency: 30 W at 11.56 GHz, 400 W at 14 GHz, 370 W at 14 GHz with additional 30 W at 11.56 GHz, and 430 W at 14 GHz only. Here the $B_{min}/B_{ECR}$ ratio was set to 0.77 (for 14 GHz) and the oxygen pressure to 3.5E-7 mbar. The only difference in the EEDs in Fig. 6 is the ratio between peaks and the total number of lost electrons, which is consistent with the aforementioned observations for single-frequency heating regime, whereas the position of the peaks in energy remains unaffected by the power / frequency combinations.

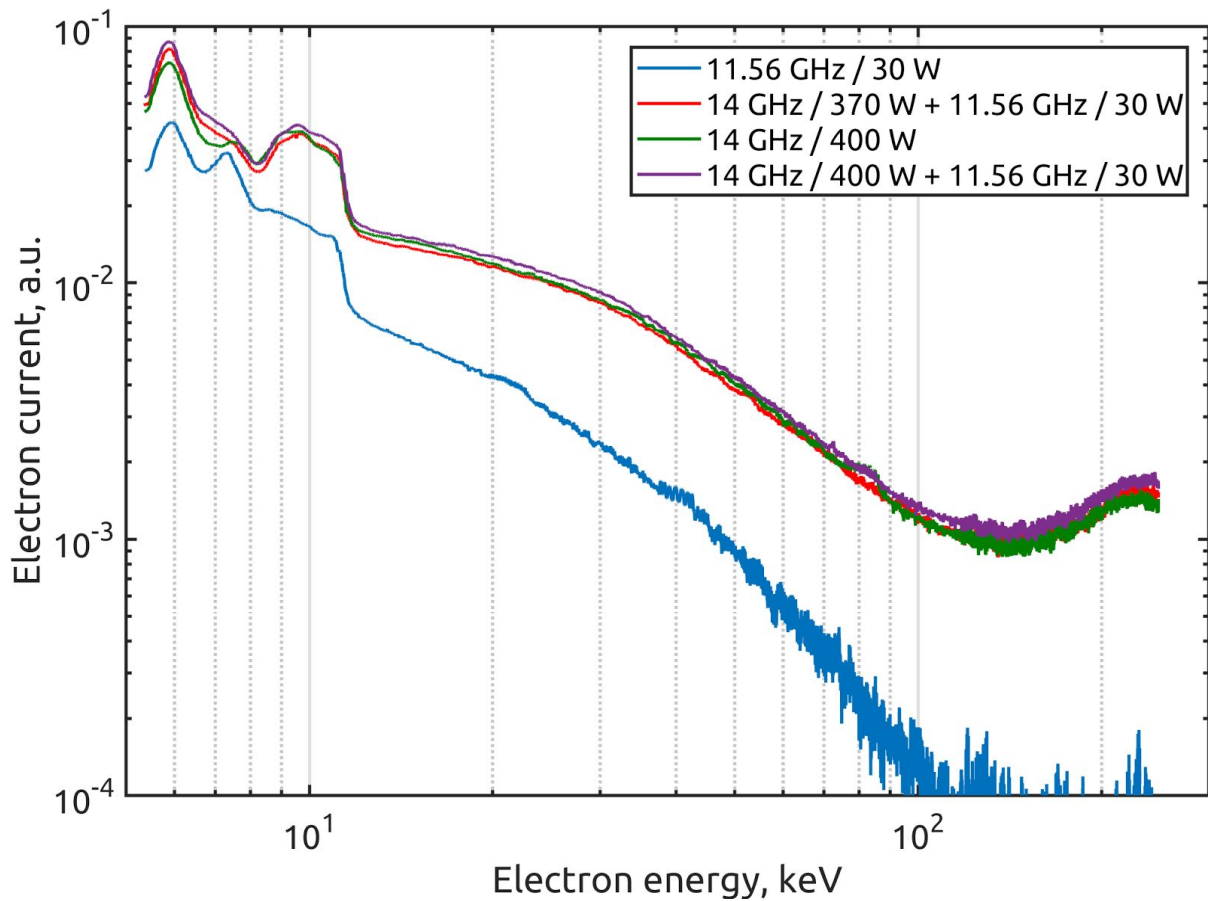

Figure 6. EEDs with the following combinations of microwave power / frequency: 30 W at 11.56 GHz, 400 W at 14 GHz, 370 W at 14 GHz + 30 W at 11.56 GHz, and 430 W at 14 GHz only. $B_{min}/B_{ECR}=0.77$ and the oxygen pressure - 3.5E-7 mbar.

**Dependence of the EED on the magnetic field strength**

The only parameter which was observed to noticeably affect the EED is the magnetic field strength. Figure 7 shows the recorded EEDs at fixed power (400 W at 14 GHz) and pressure (oxygen, 3.5E-7 mbar) but different $B_{min}/B_{ECR}$ ratios. The local maxima of the EED shift towards higher energies with increasing magnetic field strength i.e. from 5.5 keV at $B_{min}/B_{ECR}=0.77$ to 7 keV at $B_{min}/B_{ECR}=0.81$.

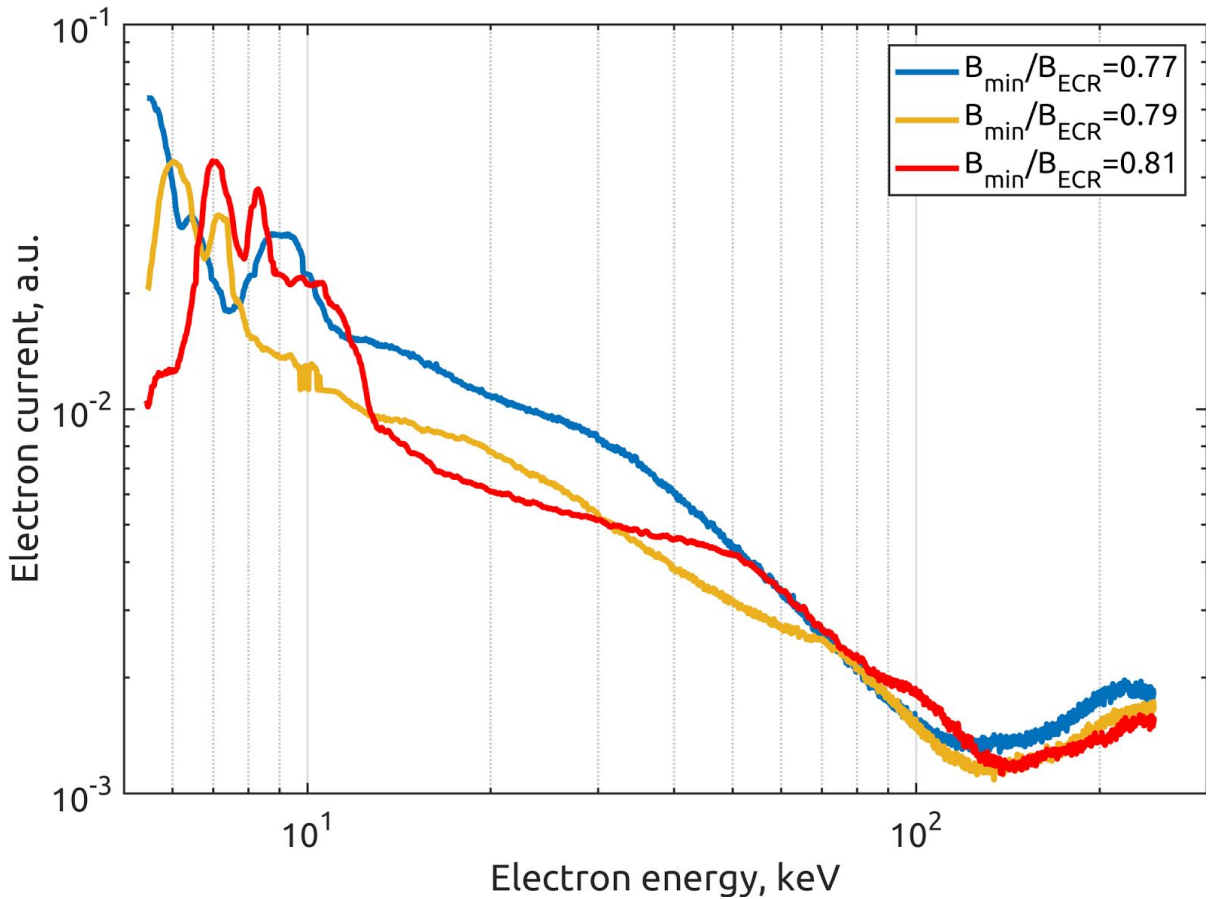

Figure 7. Dependence of the EED on the magnetic field strength at constant power (400 W at 14 GHz) and pressure (oxygen, 3.5E-7 mbar).

Appearance of plasma instabilities [26] restricted observing the shift of the EED towards higher energies at strong magnetic fields i.e. above the threshold $B_{min}/B_{ECR}$ of 0.82, in oxygen plasma. A detailed study of the EED dependence on the magnetic field strength was thus carried out with krypton plasma which is stable over a wider range of magnetic field values especially at low microwave powers. This is presumably due to increased rate of inelastic collisions in comparison to oxygen as discussed thoroughly in Ref. [27] . Figure 8 shows a density plot of the EED at logarithmic scale as a function of the magnetic field strength with 100 W of 14 GHz microwave power at krypton pressure of 3.5E-7. The scan was realized by stepping the $B_{min}/B_{ECR}$ ratio with 0.005 step and acquiring the EED at each setting. The obtained EEDs are normalized to unity and accumulated onto the density plot i.e. each horizontal line represents a single scan at given field strength. The trend of increasing energy at the peak of the distribution with the increase of magnetic field strength is clearly visible.

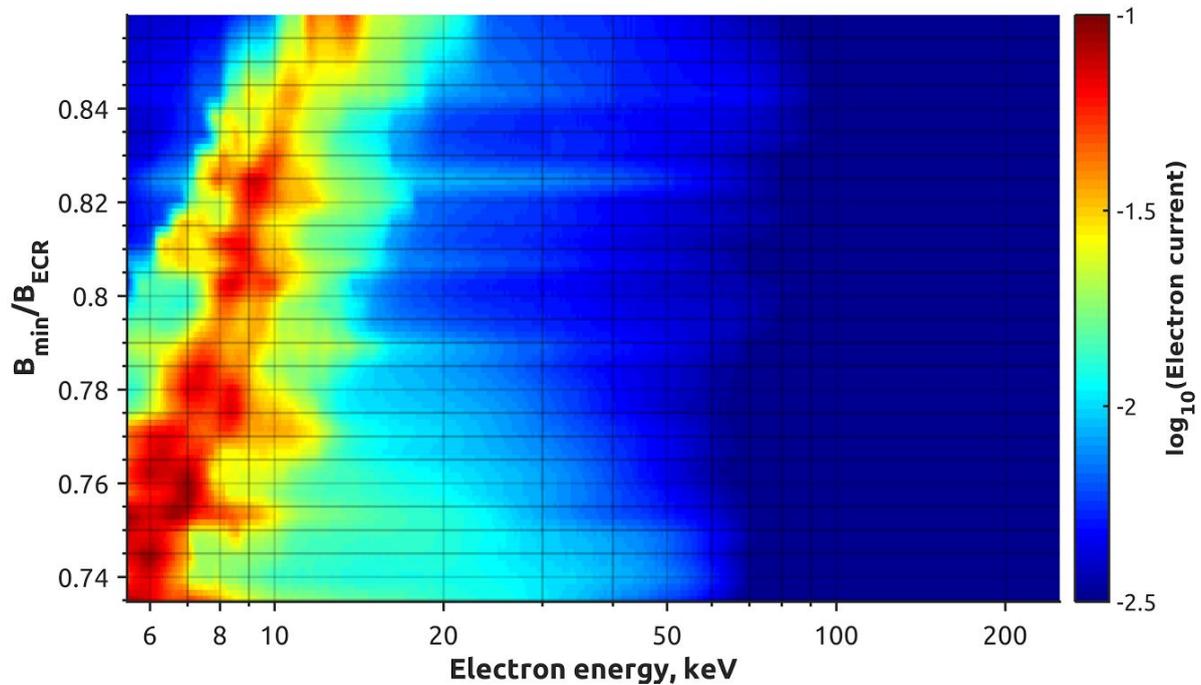

Figure 8. A density plot of the EED at logarithmic scale as a function of the magnetic field strength with 100 W of 14 GHz microwave power at krypton pressure of 3.5E-7.

Further analysis of the data in Fig. 8 yields (normalized) total number of (escaped) electrons as a function of the axial magnetic field strength, shown in Figure 9a. The increase of the magnetic field strength leads to a noticeable drop of the number of energetic electrons escaping the confinement through the extraction mirror. That might be judged as an enhancement of the electron confinement though one should keep in mind that the change of $B_{min}/B_{ECR}$ ratio may also change the spatial distribution of electron losses between the axial and radial mirrors. Figure 9b shows the dependence of the integral mean energy (average energy) calculated for each $B_{min}/B_{ECR}$ ratio. The average energy grows with the magnetic field up to $B_{min}/B_{ECR}$ ~ 0.8 and then saturates. Coincidentally, the value of $B_{min}/B_{ECR}$ ~ 0.8 is often found optimal for high charge state ion production [28].

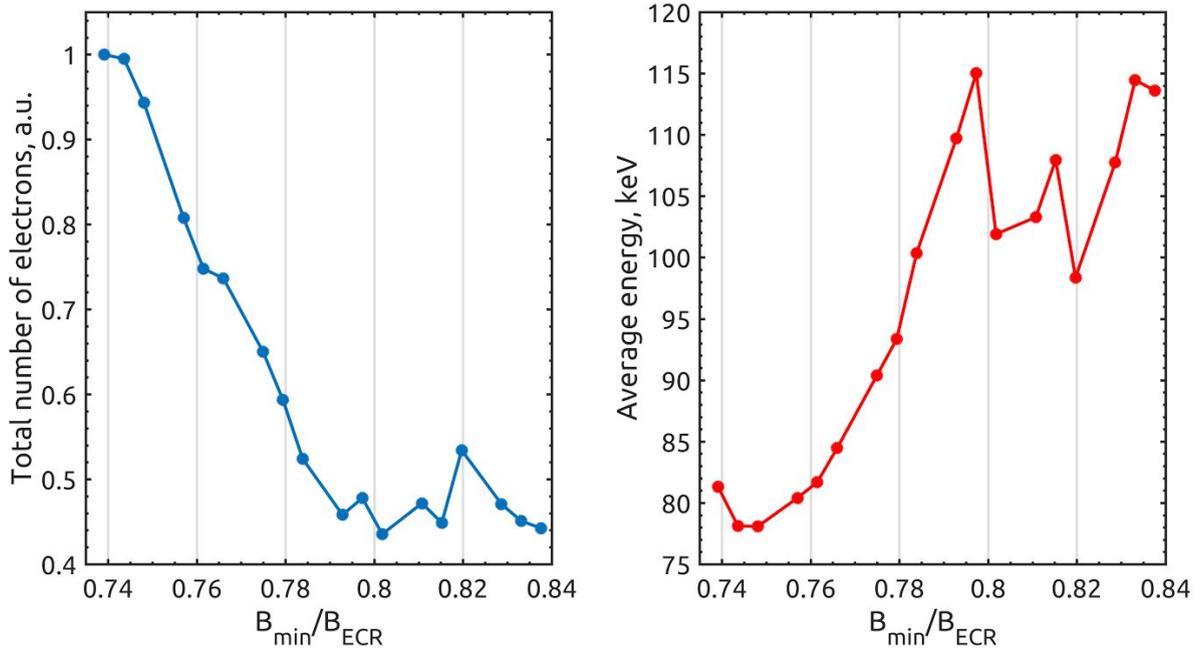

Figure 9. Total number of escaped electrons (normalized) (a) and averaged over EED electron energy (b) as a function of $B_{min}/B_{ECR}$. Heating power 100 W at 14 GHz, 3.5E-7 mbar of krypton.

## Discussion

Bremsstrahlung experiments with ECR ion sources have shown that the x-ray spectrum is most sensitive to the magnetic field, whereas other parameters (heating power and frequency, gas pressure) are less influential [14, 29]. The most recent investigation on bremsstrahlung dependence on the magnetic field strength in ECRIS has been reported in [14], showing that $B_{min}$ is basically the only parameter which affects the spectral temperature in the very similar way as it affects the average energy of lost electrons reported here (see Fig. 9b). The consistency of previous results on bremsstrahlung measurements with those presented here together with hypotheses and experimental results on RF scattering allows the use of described method for qualitative estimation of electron energy distribution function in ECRIS plasmas.

The abrupt decrease of the EEDF observed at the energies 12-18 keV for almost all experimental conditions might be correlated with a superadiabacity cut-off [8]. The electric field of the heating wave must be on the order of 50 V/cm to match with the experimental observations, which seems to be quite a low value keeping in mind the plasma chamber is a cavity with high Q-factor [30], and for an empty cavity the field value on the ECR surface was simulated to be on the order of 1 kV/cm [31]. However, a consideration of intense field damping by highly absorbing plasma might be able to remove the contradiction [32]. Further studies on this topic are required.

All of the EEDs have a linear tail at log scale (20 - 50 keV range, sometimes 30 - 100 keV), which might be mistakenly associated with a Maxwellian one. Linear fit of the described part of EEDs yields the temperature being too high for the fitting region (for some EEDs even higher than the upper fitting limit), implying that Maxwellian fit is inapplicable. Thus, it is emphasized that the EED is hardly Maxwellian.

Presented results give new insight into the process of the EEDF formation in the minimum-B confinement under ECR condition. Although the method of deconvoluting the energy distribution of lost electrons back to the energy distribution of confined electrons is a topic of future research, the suggested experimental procedure seems to be much more relevant than bremsstrahlung diagnostics in terms of estimating the efficiency of ECR heating and electron confinement in modern ECRISs, as it gives not only the "spectral temperature", but rather a fine structure of the energy distribution function (of lost electrons). Despite the fact that it is impossible to measure the EED and e.g. ion beam parameters simultaneously in contrast to bremsstrahlung, the described method is non-invasive unlike Langmuir probe diagnostics. So, it is possible to correlate plasma and /or ion beam parameters measurements with EED if the reproducibility is proven to be high, which is usually the case with modern ECRISs. However, it should be noted that the EED in case of ion beam extraction might differ from reported above, that is to be studied. In the end, it is emphasized that direct measurements of electron energies might be of interest for fundamental research in the field of ECR heating process as well as for open mirror fusion machines [33-36].

## Acknowledgements

This work was supported by the Academy of Finland under the Finnish Centre of Excellence Programme 2012-2017 (Project No. 213503) and mobility grants No. 311173 and No. 311237. The work of I. Izotov and V. Skalyga was supported by Russian Science Foundation, grant #16-12-10343.

## References

1. V. A. Zhiltsov, A. A. Skovoroda, A. V. Timofeev, K. Yu. Kharitonov, and A. G. Sherbakov, Fiz. Plazmy 17(7), 771 (1991).
2. V. A. Zhiltsov, A. Yu. Kuyanov, A. A. Skovoroda, and A. V. Timofeev, Fiz. Plazmy 20(4), 267 (1994).
3. A. Girard, C. Perret, G. Melin, and C. Lécot, Rev. Sci. Instrum. 69, 1100 (1998).
4. G. Melin, F. Bourg, P. Briand, J. Debernardi, M. Delaunay, R. Geller, B. Jacquot, P. Ludwig, T. K. N'Guyen, L. Pin, M. Pontonnier, J. C. Rocco, and F. Zadworny, Rev. Sci. Instrum. 61, 236 (1990).
5. [PSST2015-4] Barue C, Lamoreux M, Briand P, Girard A and Melin G. J. Appl. Phys. 76, 5 (1994).
6. [PSST2015-5] Douysset G, Khodja H, Girard A and Briand J P. Phys. Rev. E, 61, 3 (2000).
7. O Tarvainen et al 2014 Plasma Sources Sci. Technol. 23 025020
8. E. V. Suvorov and M. D. Tokman, Sov. J. Plasma Phys. 15, 540 (1989).
9. T Ropponen et al 2011 Plasma Sources Sci. Technol. 20 055007
10. I. V. Izotov et al., IEEE Transactions on Plasma Science, vol. 36, no. 4, pp. 1494-1501, Aug. 2008.
11. T. Thuillier et all, REVIEW OF SCIENTIFIC INSTRUMENTS 79, 02A314 (2008)

12. V. Skalyga, I. Izotov, V. Zorin, and A. Sidorov. Phys. Plasmas 19, 023509 (2012); doi: 10.1063/1.3683561
13. S Gammino et al 2009 Plasma Sources Sci. Technol. 18 045016
14. J. Benitez, C. Lyneis, L. Phair, D. Todd and D. Xie. IEEE Transactions on Plasma Science, vol. 45, no. 7, pp. 1746-1754, July 2017. doi: 10.1109/TPS.2017.2706718
15. S. Kasthurirangan, A. N. Agnihotri, C. A. Desai, and L. C. Tribedi. Review of Scientific Instruments 83, 073111 (2012); https://doi.org/10.1063/1.4738642
16. Druyvesteyn MJ (1930). "Der Niedervoltbogen". Zeitschrift für Physik. 64 (11-12): 781–798
17. J. L. Jauberteau, I. Jauberteau, O. D. Cortázar, and A. Megía-Macías. Physics of Plasmas 23, 033513 (2016); https://doi.org/10.1063/1.4944677
18. S. V. Golubev, I. V. Izotov, D. A. Mansfeld, and V. E. Semenov. Rev. Sci. Instrum. 83, 02B504 (2012); doi: 10.1063/1.3673012
19. I. Izotov, D. Mansfeld, V. Skalyga, V. Zorin, T. Grahn et al. Phys. Plasmas 19, 122501 (2012); doi: 10.1063/1.4769260
20. V Toivanen, T Kalvas, H Koivisto, J Komppula and O Tarvainen. 2013 JINST 8 P05003
21. Yinghong Lin and David C. Joy. Surf. Interface Anal. 2005; 37: 895–900
22. http://rcwww.kek.jp/egsconf/proceedings/proc14_errata/01-tabata.pdf
23. O. Tarvainen, P. Suominen, T. Ropponen, T. Kalvas, P. Heikkinen, and H. Koivisto, REVIEW OF SCIENTIFIC INSTRUMENTS 76, 093304 (2005).
24. L. Celona et al., Rev. Sci. Instrum. 79, 023305 (2008).
25. Z. Q. Xie and C. M. Lyneis, in Proceeding of 12th International Workshop on ECR Ion Sources, Riken (RIKEN, Institute for Nuclear Study, Japan, 1995), p. 24.
26. O Tarvainen et al 2014 Plasma Sources Sci. Technol. 23 025020
27. ECRIS14_Tarvainen
28. D. Hitz, A. Girard, G. Melin, S. Gammino, G. Ciavola and L. Celona. Rev. Sci. Instrum. 73, (2002), p. 509.
29. D. Leitner, J. Y. Benitez, C. M Lyneis, D.S. Todd, T. Ropponen, J. Ropponen, H. Koivisto and S. Gammino, Rev. Sci. Instrum. 79, 033302 (2008).
30. V. Toivanen, O. Tarvainen, C. Lyneis, J. Kauppinen, J. Komppula, and H. Koivisto. Review of Scientific Instruments 83, 02A306 (2012); https://doi.org/10.1063/1.3660818
31. O. Tarvainen, J. Orpana, R. Kronholm, T. Kalvas, J. Laulainen, H. Koivisto, I. Izotov, V. Skalyga, and V. Toivanen. Review of Scientific Instruments 87, 093301 (2016); doi: 10.1063/1.4962026
32. T. Ropponen, O. Tarvainen, P. Suominen, T.K. Koponen, T. Kalvas, H. Koivisto. Nuclear Instruments and Methods in Physics Research A 587 (2008) 115–124. https://doi.org/10.1016/j.nima.2007.12.030
33. D.V. Yakovlev, A.G. Shalashov, E.D. Gospodchikov, A.L. Solomakhin, V.Ya. Savkin and P.A. Bagryansky, Electron cyclotron plasma startup in the GDT experiment // Nucl. Fusion 57 (2017) 016033.
34. P.A. Bagryansky, E.D. Gospodchikov, Yu.V. Kovalenko, A.A. Lizunov, V.V. Maximov, S.V. Murakhtin, E.I. Pinzhenin, V.V. Prikhodko, V.Ya. Savkin, A.G. Shalashov, E.I. Soldatkina, A.L. Solomakhin, D.V. Yakovlev. Electron Cyclotron Resonance Heating Experiment in the GDT Magnetic Mirror: Recent Experiments and Future Plans // Fusion Science and Technology, Vol. 68, Issue 1, Pages 87-91 (2015).

35. P.A. Bagryansky, A.V. Anikeev, G.G. Denisov, E.D. Gospodchikov, A.A. Ivanov, A.A. Lizunov, Yu.V. Kovalenko, V.I. Malygin, V.V. Maximov, O.A. Korobeinikova, S.V. Murakhtin, E.I. Pinzhenin, V.V. Prikhodko, V.Ya. Savkin, A.G. Shalashov, O.B. Smolyakova, E.I. Soldatkina, A.L. Solomakhin, D.V. Yakovlev, K.V. Zaytsev, Overview of ECR plasma heating experiment in the GDT magnetic mirror // Nucl. Fusion 55 (2015).
36. Ceccherini et al. 2014, APS Meeting Abstracts.